# Pressure-Induced Structure Transitions in Eu Metal to 92 GPa

W. Bi,[1] Y. Meng,[2] R. S. Kumar,[3] A. L. Cornelius,[3] W. W. Tipton,[4] R. G. Hennig,[4] Y. Zhang,[3] C. Chen,[3] and J. S. Schilling[1]

[1]Department of Physics, Washington University, CB 1105, One Brookings Dr, St. Louis, Missouri 63130, USA
[2]HPCAT, Carnegie Institution of Washington, 9700 S. Cass Ave., Argonne, Illinois 60439, USA
[3]Department of Physics and High Pressure Science and Engineering Center, University of Nevada, Las Vegas, Nevada 89154, USA
[4]Department of Materials Science and Engineering, Cornell University, Ithaca, New York 14853, USA

Synchrotron x-ray diffraction experiments have been carried out on Eu metal at ambient temperature to pressures as high as 92 GPa (0.92 Mbar). Following the well-known bcc-to-hcp transition at 12 GPa, a mixed phase region is observed from 18 to 66 GPa until finally a single orthorhombic (*Pnma*) phase persists from 66 to 92 GPa. These results are compared to predictions from density functional theory calculations. Under pressure the relatively large molar volume $V_{mol}$ of divalent Eu is rapidly diminished, equaling or falling below $V_{mol}(P)$ for neighboring trivalent lanthanides above 15 GPa. The present results suggest that above 15 GPa Eu is neither divalent nor fully trivalent to pressures as high as 92 GPa.

## I. Introduction

In contrast to the other lanthanide metals, which are trivalent, Eu and Yb retain the divalency of their free-atom state. As a result, their atomic volumes are significantly larger and their structures do not fit into the normal structure sequence across the trivalent lanthanide series (hcp→Sm-type→double hcp→fcc→distorted fcc) either with increasing pressure or decreasing atomic number [1-3]. At pressures near 1 Mbar, Yb takes on the hexagonal *hP*3 structure exhibited by the light actinides Sm and Nd under pressure, providing evidence that for pressures of 1 Mbar and above Yb has joined the regular lanthanide series and become fully trivalent [4]. The equation of state (EOS) of Yb is consistent with this conclusion [4].

Structure studies on Eu metal at ambient temperatures, on the other hand, have only been carried out to 43 GPa [5, 6], revealing a bcc-to-hcp transition at 12 GPa accompanied by a 4% volume collapse, with a new close-packed structure appearing near 17 GPa. From the fact that the EOS of Eu approaches that of trivalent Gd near 20 GPa, it was concluded that at this pressure a significant increase in Eu's valence must have occurred [5]. $L_{III}$ absorption [7] and Mössbauer effect [8, 9] studies reportedly indicate that at 10 GPa Eu's valence has already increased to approximately 2.5, with a further increase to 2.64 at pressures of 34 GPa. Theoretical predictions of the pressure necessary for the full divalent-to-trivalent transition in Eu vary from 35 GPa [2, 10] to 71 GPa [11].

Should sufficiently high pressure be applied to bring Eu to full trivalency $Eu^{3+}$ ($4f^6$ where J = 0), its magnetic ground state $Eu^{2+}$ ($4f^7$ where J = 7/2) would be destroyed, leaving only weak Van Vleck paramagnetism which can coexist with superconductivity. Indeed, trivalent $Am^{3+}$ ($5f^6$) is a Van Vleck paramagnet which superconducts below 0.79 K [13]. Since the other trivalent *s,p,d*-electron metals, Y, Sc, La and Lu all superconduct at temperatures 10 – 20 K at 1 Mbar pressure [12], one would anticipate that trivalent Eu superconducts at comparable temperatures. Eu was recently found to become superconducting for pressures higher than 80 GPa [14]. However, we note that the value of its superconducting transition temperature $T_c \approx 2$ K and pressure derivative $dT_c/dP \approx +0.018$ K/GPa are both much less than those reported for the trivalent *s,p,d*-metals Sc, Y La, and Lu [12]. This was taken to indicate that to 142 GPa Eu does not become fully trivalent, but rather mixed valent [14]. Other possibilities are that the Van Vleck paramagnetism of trivalent Eu weakens the superconducting state or that the crystal structure taken on by Eu in the pressure range of 80 - 142 GPa is not favorable for higher values of $T_c$ [14].

Extending the previous x-ray diffraction studies on Eu metal to pressures above 80 GPa is important for several reasons: (1) to establish whether the sudden appearance of superconductivity near 80 GPa is associated with a structural phase transition, (2) to check whether Eu's EOS does indeed approach that of trivalent Gd near 20 GPa as reported earlier [5], (3) to test for pressure-induced trivalency in Eu by establishing whether at extreme pressures the structures taken on by Eu follow those of the regular trivalent lanthanide series.

In the present experiments on Eu metal to 92 GPa, three structure phase regions are observed: a bcc-to-hcp transition near 12 GPa, a mixed phase region from 18 to 62 GPa, and then a transition to a single-phase orthorhombic (*Pnma*) structure at 66 GPa which is retained to the highest pressure applied 92 GPa. This pressure-induced structure sequence is compared to the results of a theoretical calculations based on density function theory (DFT). The present results suggest that above 15 GPa Eu is neither divalent nor fully trivalent to pressures as high as 92 GPa.

**II. Experimental Techniques**

High-pressure synchrotron angle-dispersive x-ray diffraction experiments were performed at beamline 16ID-B of the High Pressure Collaborative Access Team (HPCAT) at the Advanced Photon Source (APS), Argonne National Laboratory. A symmetric cell was used with 1/6-carat, type Ia diamond anvils with 0.18 mm culets beveled at 7 degrees out to 0.35 mm. The Re gaskets were pre-indented from the original thickness of 250 micron to 30 micron central thickness; a 60 μm dia. hole was electro-spark drilled through the center to form a sample chamber. The high-purity Eu sample (99.98% metals basis), obtained from the Materials Preparation Center of the Ames Laboratory [15], was loaded into the sample chamber in an Ar glove box due to the high reactivity of the sample. A small amount of Pt powder (~20 μm) was placed on the sample as a pressure marker [16].

The monochromatic x-ray beam (29.879 keV, 34.221 keV and 29.130 keV) used in three separate experiments was focused to less than 10 μm at the sample location in both horizontal and vertical directions. Due to the soft nature of the sample, the pressure difference between the center and edge of the sample chamber is only 1 GPa at 88 GPa, allowing us to determine the critical pressure for a given phase transition quite accurately. Normally, the diffraction pattern shows only peaks from the sample and the Pt marker. However, for pressures of 55 GPa and above, weak peaks from the Re gasket were observed in some measurements due to the irregular shape of the gasket hole. Diffraction patterns were collected at room temperature and high pressures using an image plate detector (MAR345) with an exposure time of typically 2 to 15 s. The sample-to-detector distance was precisely calibrated using a NIST $CeO_2$ standard. Fig. 1 shows x-ray diffraction images of Eu at pressures of 4, 14 and 92 GPa. In order to be consistent with the superconductivity experiments [14], no pressure medium was used in the present studies. The x-ray diffraction peaks became quite broad at the highest pressures (see the data at 92 GPa), presumably due to sizeable strains in the sample from the non-hydrostatic pressure environment. The results of these diffraction experiments are discussed in detail in Section IV.

## III. Results from Density Functional Theory Calculations
### A. Structure Search Methods

Modern structure search methods can efficiently determine the crystal structure of materials under extreme conditions, such as high pressures, which are challenging to reach experimentally. These methods can effectively overcome kinetic barriers to the formation of ground-state structures and cover vast ranges of composition and pressure which can be expensive and time consuming to explore *in situ*.

Computational structure search methods optimize structural parameters to minimize the Gibbs free energy to obtain thermodynamic ground-state structures. At low temperatures, the Gibbs free energy is commonly approximated by the enthalpy, which may be calculated within density functional theory (DFT). Crystal structure searches present a particularly challenging optimization problem since the space of possible solutions is large and the objective function is poorly understood and expensive to compute. Therefore, stochastic search and heuristic algorithms are generally used to attack this problem.

The simplest stochastic search method is the random search. A crystal lattice and atomic locations are generated randomly subject to constraints such as maximum and minimum lattice parameters and inter-atomic distances. These trial structures are then relaxed and the enthalpy is evaluated using DFT to determine the ground state structure at a given pressure. This algorithm has the advantages of being conceptually simple and quick to program and is often quite successful in the study of simple systems such as those of a single element. However, the algorithm is not very efficient as it does not make use of the information which is generated about the system as it progresses.

Heuristic methods such as genetic algorithms attempt to address this issue. Genetic algorithms create a sequence of collections or "generations" of trial structures. The first generation is generated randomly as in the random search method. Applying various biologically inspired operators to the previous generation creates subsequent generations. These operators select one or more "parent" structures and attempt to combine them into a "child" solution, which maintains the favorable aspects of its parents. Structures with lower energy are more likely to be chosen as parents. In this way, structural features which lead to low energies are propagated in the population. This method is significantly more efficient than the random search in solving complicated materials systems.

We have implemented a random search algorithm and an unpublished genetic algorithm to search for novel Eu structures under pressure. Additionally, the search for stable and most competitive metastable structures of Eu under pressure has been performed by the evolutionary algorithm code USPEX [17].

### B. Density Functional Method

The structure relaxations and enthalpy calculations are performed using VASP (Vienna *ab initio* simulation program) employing the projector augmented wave (PAW) method within the frozen-core approximation [18]. The PAW potential describes the [Kr] $5s^2\ 4d^{10}\ 4f^7$ states as core states, neglecting the effect of $f$ electrons on the bonding. The approximation of localized $f$ electrons is expected to be accurate at lower pressures but will need to be checked in future work for the high-pressure structures. For the bcc Eu structure, neglecting the $f$ electron effects only changes the lattice parameters from 4.46 Å to 4.44 Å for a ferromagnetic ordering of the $f$ electrons. Note that below 91 K the $f$ electrons in bcc Eu order magnetically in an incommensurate spin spiral structure [19] with the period of $3.6a$, where $a$ is the lattice parameter, which is beyond the scope of our structure search.

For the random search and in-house genetic algorithm code, the generalized gradient approximation of Perdew, Becke and Ernzerhof (PBE) is used [20]. A cutoff energy of 400 eV and a $k$-point mesh with a density of 50 Å$^{-1}$ ensure convergence of the total energy to 1 meV/atom.

The structures predicted by USPEX are fully relaxed using the PW91 exchange-correlation functional [18]. For these calculations, a sampling of at least 80 nonequivalent $k$-points is used during the structure search, resulting in a total energy convergence of better than 2 meV/atom. More accurate DFT calculations were performed to obtain the energies of stable and metastable structures. For these, an energy cutoff of 400 eV and a 12x12x12 $k$-point mesh are used to obtain better convergence on total energy and stress.

**C. Results of Structure Search**

We performed structure searches with the random search method and in-house evolutionary algorithm at pressures of 0, 20, 40, 60 and 80 GPa for structures with up to 8 atoms per unit cell for the random search method and up to 30 atoms per unit cell for the evolutionary algorithm. The search discovered a large number of candidate crystal phases with low enthalpies (bcc, fcc, hcp, *Fdd*2, *Pnma*, *Fddd*, *Cc*, *Imm*2, *R-3m*, *C2/m* and *C2/c*). All of these structures have enthalpies within a range of 50 meV/atom, and we expect that the enthalpic ordering of the structures and the transition pressures could be affected by the localization of the $f$ electrons in our calculations.

We calculate the enthalpy as a function of pressure for all of these trial structures. Fig. 2 shows the enthalpies of the predicted ground state structures and their stability ranges. We find the bcc phase at low pressures and a transition to the hcp structure at 10 GPa. At a pressure of 16 GPa we predict a transformation to the *C2/c* structure, at 22 GPa to the *Fdd*2 structure, and at 34 GPa to the *Pnma* structure. The *Pnma* structure is nearly degenerate to the *C2/c* structure, and we predict that the *C2/c* phase is slightly lower in enthalpy above 46 GPa. However, these enthalpy differences are below the accuracy limits of current approximations of the exchange-correlation functional in DFT calculations [21].

The USPEX structure predictions are based on calculations using unit cells containing four Eu atoms. No pre-imposed symmetry constraint or experimental information is used in the calculations. The first generation of structures is generated randomly and there are up to 30 structures in each generation. The most favorable 65% structures of each generation are chosen to predict the next generation by heredity (60%), mutation (20%) and permutation (20%). The USPEX algorithm successfully finds the known bcc and hcp Eu at ambient pressure and at 15 GPa, respectively. It is then used to search for structures at higher pressures. We carried out USPEX searches at 25, 30, 45, 70, and 90 GPa. Up to 25 generations of candidates are produced to find the lowest-enthalpy structure. As shown in Fig 3, these calculations find that orthorhombic Eu with the space group *Pnma* (No. 62) is stable from 25 to 70 GPa. At a pressure of 90 GPa, hcp Eu with the space group $P6_3/mmc$ (No. 194) reappears. The calculated enthalpies as a function of pressure for the hcp and *Pnma* structure relative to that of bcc Eu are plotted in Fig. 3. They show that bcc Eu would be expected to transform to hcp at about 10 GPa and then to *Pnma* above 20 GPa. In the calculations we use the hcp structure with four-atom orthorhombic unit cell (Fig. 3(d)) by applying the transformation a′ = 2a + b, b′ = b, c′ = c, where a, b and c are lattice vectors of the primitive hcp cell (Fig. 1(b)). The corresponding atomic position is (1/6, 1/2, 1/4) and the ratio of |a′|/|b′| is $\sqrt{3}$. For the *Pnma* structure, the typical atomic position has the form {(1/6, -x + 1/2, 1/4) 0 < x < 0.1} and the ratio |a′|/|b′| deviates from that of hcp structure. Therefore, the *Pnma* structure can be viewed as a distorted hcp structure in the orthorhombic cell. However, new intermediate structures with more than four atoms per unit cell could still be possible in this pressure region. The hcp structure has lower enthalpies than the *Pnma* structure above 80 GPa.

Discrepancies in the findings of the two genetic algorithms are due to constraints placed on the search space. The USPEX search was constrained to 4-atom unit cells whereas the in-house code considered structures with up to 30 atoms per unit cell. Indeed, the stable structures found by USPEX are a subset of those found by the other search, and the two that it missed, *C2/c* and *Fdd2*, both have unit cells of greater than 4 atoms.

In the pressure-induced superconducting state in Eu metal, the underlying electron pairings could be mediated by lattice vibrations (BCS framework) as for Sc, Y, La, and Lu [12]. It is, therefore, interesting to calculate the lattice dynamics of Eu at high pressures. To this end we have performed the phonon density of states (PDOS) calculations using the direct-force method as implemented in the PHONON package. This method has been proven reliable in studying bcc Eu at ambient pressure [19]. We first study the PDOS of bcc and hcp Eu at 0 and 15 GPa as shown in Fig. 4(a). The obtained phonon peaks are consistent with previous calculation and experiment [9, 19]. We then calculated the PDOS of both hcp and *Pnma* phase at 90 GPa (Fig. 4(b)). The similarly positioned major peaks at 9 and 30 meV for both phases are ascribed to their close structural features. It is interesting to note that a major low-frequency PDOS peak at 9 meV is present in both the lower pressure (15 GPa) and higher pressure (90 GPa) hcp phases. Further evaluations of the electron-phonon coupling strength, as well as parallel work on La and Lu, are needed to gain a detailed understanding of superconductivity in Eu at high pressures.

We note that Nixon and Papaconstantopoulos [22] have recently calculated the electronic structure of Eu for the bcc, hcp, and fcc structures to 90 GPa pressure using the augmented-plane-wave method in the local-density approximation. Using a simple Debye model to approximate the change in the average phonon frequency under pressure, they find that in both the bcc and hcp phases Eu becomes superconducting above 60 GPa, increasing to a value near 2 K at 80 GPa, in agreement with experiment [14].

**IV. Results of Experiment**

In our data analysis the two-dimensional images (see Fig. 1) were integrated to give intensity as a function of diffraction angle (2θ) using the software FIT2D [23]. The Le Bail and Rietveld refinements were performed using LHPM-RIETICA [24] and GSAS [25]. Three separate high-pressure experiments were carried out. In the first run, XRD data were collected at pressures from 4 to 43 GPa; a gasket failure prevented measurements to higher pressures. In the other two runs, the highest pressure reached was 92 GPa. In all three experiments, diffraction images were collected at 2-5 GPa intervals with increasing and decreasing pressure. The observed phase transition pressures in these experiments are consistent with each other.

Typical x-ray diffraction spectra for Eu metal at four pressures to 35 GPa, including the results of a full-profile Rietveld refinement for bcc (*Im-3m*) and hcp (*P6₃/mmc*), are shown in Fig. 5. Since Pt was used as an internal pressure standard, its fcc (*Fm-3m*) phase is included as a second phase in the refinement. The anticipated abrupt phase transition in Eu from bcc to hcp near 12 GPa is clearly observed. Above 18 GPa several new peaks begin to appear as shown with arrows in Fig. 5, indicating a sluggish phase transition. This result is consistent with studies by Takemura and Syassen [5] where silicone oil served as pressure medium. In the present experiments the spectra up to 28 GPa can still be indexed with hcp if the weak peaks are excluded. Above 30 GPa the phase change proceeds more rapidly. The anomalies at 18 and 28 GPa observed by Bundy and Dunn [26] in the room-temperature electrical resistivity are possibly related to these changes in structure.

In our first attempt to solve the post-hcp phase we tried a multiple hcp cell with 36 atoms ($a' = a$, $c' = 18c$, $a$ and $c$ are the cell parameters from previous hcp phase), as proposed by Takemura and Syassen [5]. However, such a multiple hcp cell, which is quite unusual, would yield a large number of diffraction peaks not observed in the present experimental data. We also considered the orthorhombic *Fdd*2 space group suggested by the theoretical prediction shown in Fig. 2. However, *Fdd*2 has a large unit cell with 40 atoms and has high-intensity peaks at angles lower than the first peak observed in our experiments, irrespective of the actual atomic positions. Even though the indexing of the XRD patterns above 18 GPa show agreement with both *P*1 and *C*2/*c* space groups, the *C*2/*c* space group seems more likely as the symmetry is higher and also the theoretical calculations above 18 GPa find the enthalpy of the monoclinic *C*2/*c* structure to be the lowest among the candidates examined (see Fig. 2). Hence, this phase is assigned as the post hcp phase and further refinements were carried out in the mixed-phase region between 18 and 62 GPa.

As pressure is increased to 41 GPa, an orthorhombic phase *Pnma* coexisting with *C*2/*c* appears (Fig. 6). The refinement of the mixed phase at 55 GPa is shown in Fig. 6 including the Le Bail fit of *C*2/*c* with cell parameters $a = 3.134(3)$ Å, $b = 4.970(7)$ Å, $c = 9.301(5)$ Å, $\beta = 106.65(10)°$ and Rietveld fit of *Pnma* with cell parameters $a = 5.042(2)$ Å, $b = 4.357(2)$ Å, $c = 3.023(1)$ Å with Eu on 4c sites and $x = 0.327(1)$, $y = 1/4$, $z = 0.035(1)$. From 41 to 92 GPa, the two peaks (see arrows in Fig. 6) belonging to the *C*2/*c* phase, which cannot be indexed with *Pnma*, gradually merge into the next peak at higher angle. Above 66 GPa both peaks have vanished and the spectra can be indexed as single phase *Pnma*. The cell parameters of *Pnma* at 75 GPa are $a = 4.977(1)$ Å, $b = 4.264(1)$ Å, $c = 2.944(1)$ Å and the Rietveld refinement is shown in Fig. 6.

Table I summarizes the cell and atomic position parameters and refinement residues for bcc, hcp, and orthorhombic *Pnma* at selected pressures. Since the phase transition from hcp to *C*2/*c* and then to *Pnma* is sluggish and continuous, we are unable to determine the detailed atomic arrangement for *C*2/*c*.

## V. Discussion

The lattice parameters and their ratios are plotted under pressure to 92 GPa in Fig. 7. Between 12 and 35 GPa the parameters are obtained based on the peaks from the hcp phase (*P6₃/mmc*), while above 35 GPa the peaks from orthorhombic (*Pnma*) are used. The agreement with the lattice parameters from Ref. [5] is reasonable. As seen in Fig. 7(b), the $c/a$ ratio shows a slope change near 18 GPa when Eu enters a mixed phase. The change in the slope of $c/a$ versus pressure may signal a pressure-induced magnetic transition [27, 28].

In Fig. 8 the relative volume $V/V_o$ of Eu metal, where $V_o$ is the molar volume at ambient pressure, is plotted versus pressure and compared to previous results [5, 6, 29]. For pressures between 18 and 37 GPa, where Eu shows a mixed phase of hcp and *C*2/*c*, the volume is calculated from the hcp structure and at higher pressures from the *Pnma* structure. Corresponding to the change in slope of *c/a* in Fig. 7, there is a slight anomaly in *V(P)* near 18 GPa. All $V/V_o$ data are tabulated in Table II. The volume-pressure dependence found in the present study is seen to be in good agreement with the previous static data to 43 GPa [6]. The fit of the *V*(P) data in the bcc phase to 12 GPa using the third order Birch-Murnaghan equation [30] yields the bulk modulus $B_o$ = 10.9(6) GPa and the pressure derivative $B'_o$ = 3.0(2), both of which are close to published values [5, 6, 29].

As seen in Fig. 8, the present *V*(P) data for both increasing or decreasing pressure agree reasonably well. The pressure at which a given phase transition occurs agrees within 1-2 GPa for all three runs for both increasing and decreasing pressure. The relative volume jump at the bcc-to-hcp transition at 12 GPa is ~ 3% which is comparable to the value 4% reported in Ref. 5. No measurable volume discontinuity is observed for the phase transitions at higher pressures.

In Fig. 9 the molar volume of Eu is plotted versus pressure to 100 GPa (1 Mbar) and compared to data on the neighboring trivalent lanthanides Nd [31], Sm [32], Gd [33], and Tb [34]. The calculated molar volume for Eu in the divalent state to 42 GPa from Johansson and Rosengren is also shown (dashed line). From this figure the molar volume of Eu is seen to initially decrease rapidly under pressure from its large divalent value, falling somewhat below that for trivalent Gd at 10-20 GPa, in agreement with earlier studies by Takemura and Syassen [5]. As these authors point out, this suggests that Eu is no longer divalent above 10-20 GPa, but rather mixed valent or possibly trivalent. A fully trivalent Eu state, however, does not appear likely to pressures as high as 92 GPa since none of the structures observed under pressure in the present experiment (bcc→hcp→mixed phase→primitive orthorhombic (*Pnma*)) have been observed to pressures exceeding 1 Mbar for the neighboring trivalent rare earth metals Nd [31], Sm [32], Gd [33] and Tb [34] or, for that matter, for *any* of the other trivalent rare-earth metals. At 90 GPa Nd, Gd, and Tb all take on the monoclinic structure *C*2/*m*. As pointed out by Takemura and Syassen [5], the fact that Eu's molar volume falls below that of trivalent Gd and Tb above 20 GPa does not necessarily imply that Eu is trivalent, but rather may arise from a mixing entropy term in a mixed valent state, as treated in the valence fluctuation model of Wohlleben [35].

In Fig. 2 it is seen that the calculated enthalpies for the high-pressure phases *C*2/*c*, *Fdd*2 and *Pnma* lie quite close together between 16 and 45 GPa so that the energetically favorable phase might not be able to form under the conditions of the present experiment. Besides the enthalpy barriers, the stress anisotropies in the present non-hydrostatic pressure experiment may also affect the value of the transition pressure and the phases assumed by Eu. The *C*2/*c* phase predicted in Fig. 2 for the pressure range 46 - 80 GPa thus might not appear in experiment. Both DFT calculations predict that the orthorhombic *Pnma* phase should transform into the hcp phase above 80 GPa. However, no further phase transition was observed after *Pnma* in the present experiment to 92 GPa. An extension of these synchrotron x-ray diffraction studies to higher pressures would test this prediction of both theories.

In summary, we have carried out x-ray diffraction experiments in a DAC on polycrystalline Eu metal under pressures to 92 GPa and have observed three pressure-induced phase transitions from bcc to hcp to a mixed phase and then to an orthorhombic *Pnma* phase above 66 GPa. That Eu's large molar volume reduces to a value below that of Gd for 10 - 20 GPa pressure would appear to indicate that Eu is no longer divalent at or above these pressures. Although the equation of state does not permit a reliable estimate of Eu's valence at 92 GPa, the fact that the crystal structures assumed differ from those exhibited by Eu's trivalent neighbors at similar pressures gives evidence that Eu does not reach full trivalency at 92 GPa. Future synchrotron x-ray diffraction and spectroscopy (XANES, XMCD, Mössbauer effect) experiments on Eu metal (preferably in single crystalline form) to higher, nearly hydrostatic pressures would be desirable to more fully characterize Eu's structural phase diagram, valence state and magnetic properties, and to help understand the effect of hydrostaticity on the phase transformation kinetics and the atomic position parameters of different high pressure polymorphs.

**Acknowledgements.** Thanks are due to R. W. McCallum and K. W. Dennis of the Materials Preparation Center, Ames Laboratory, for providing the high-purity Eu sample. Work at both Washington University and the Advanced Photon Source (APS) was supported by the Carnegie/DOE Alliance Center (CDAC) through NNSA/DOE grant number DE-FC52-08NA28554, the Center for Materials Innovation, and the National Science Foundation through grant number DMR-0703896. HPCAT is supported by CIW, CDAC, UNLV and LLNL through funding from DOE-NNSA, DOE-BES and NSF. APS is supported by DOE-BES, under Contract No. DE-AC02-06CH11357. The UNLV High Pressure Science and Engineering Center was supported by the U.S. Department of Energy, National Nuclear Security Administration, under Co-operative agreement number DE-FC52-06NA26274. The work at Cornell was supported by the U.S. Department of Energy under Contract No. DE-FG05-08OR23339 and the National Science Foundation under Contract No. EAR-0703226. W.W.T. acknowledges support from the NSF IGERT fellowship under Contract No. DGE-0903653. This research used computational resources of the National Center for Supercomputing Applications under Grant No. DMR050036 and the Computational Center for Nanotechnology Innovation at Rensselaer Polytechnic Institute.

**Table I.** Cell and atomic position parameters and refinement residues of Eu in bcc, hcp and orthorhombic structure at room temperature. Calculated lattice parameters keeping the *f*-electrons in the core are given in square brackets. The deviations between the experimental and computed lattice parameters increase with pressure, reaching values up to about 6%.

|  | P = 4 GPa<br>*Im*-3*m* (bcc) | P = 14 GPa<br>*P*6$_3$/*mmc* (hcp) | P = 75 GPa<br>*Pnma* (orthorhombic) |
|---|---|---|---|
| Cell parameters (Å) | a = 4.1961 (1) [4.14] | a = 3.3501(1) [3.32]<br>c = 5.2962(2) [5.00] | a = 4.977(1) [4.77]<br>b = 4.264(1) [4.13]<br>c = 2.944(1) [2.78] |
| Atomic position parameters | x = 0<br>y = 0<br>z = 0 | x = 1/3<br>y = 2/3<br>z = 1/4 | x = 0.325(1) [0.33]<br>y = 1/4<br>z = 0.029(2) [0.08] |
| Refinement residue ($R_{wp}$) (%) | 5.4 | 5.6 | 5.8 |

**Table II.** V/V$_o$ data of Eu for increasing pressure to 92 GPa. The ambient pressure molar volume $V_o$ = 28.98 cm$^3$/mol [7] is used.

| bcc (Im-3m) | | hcp (P6$_3$/mmc) | | orthorhombic (Pnma) | |
|---|---|---|---|---|---|
| P (GPa) | V/V$_o$ | P (GPa) | V/V$_o$ | P (GPa) | V/V$_o$ |
| 0 | 1 | 12.0 | 0.564 | 40.0 | 0.386 |
| 4.0 | 0.771 | 14.0 | 0.537 | 41.5 | 0.383 |
| 5.3 | 0.734 | 15.8 | 0.519 | 43.0 | 0.380 |
| 7.8 | 0.669 | 17.0 | 0.503 | 44.5 | 0.375 |
| 9.0 | 0.631 | 19.0 | 0.481 | 48.0 | 0.367 |
| 10.6 | 0.603 | 20.8 | 0.465 | 51.0 | 0.363 |
| | | 23.0 | 0.452 | 55.0 | 0.358 |
| | | 25.0 | 0.441 | 59.0 | 0.352 |
| | | 27.0 | 0.430 | 62.0 | 0.346 |
| | | 28.0 | 0.420 | 66.0 | 0.344 |
| | | 30.4 | 0.413 | 69.0 | 0.338 |
| | | 31.0 | 0.410 | 72.0 | 0.331 |
| | | 33.0 | 0.404 | 75.0 | 0.325 |
| | | 35.0 | 0.396 | 85.5 | 0.315 |
| | | 37.0 | 0.391 | 88.0 | 0.312 |
| | | | | 90.0 | 0.310 |
| | | | | 92.0 | 0.308 |

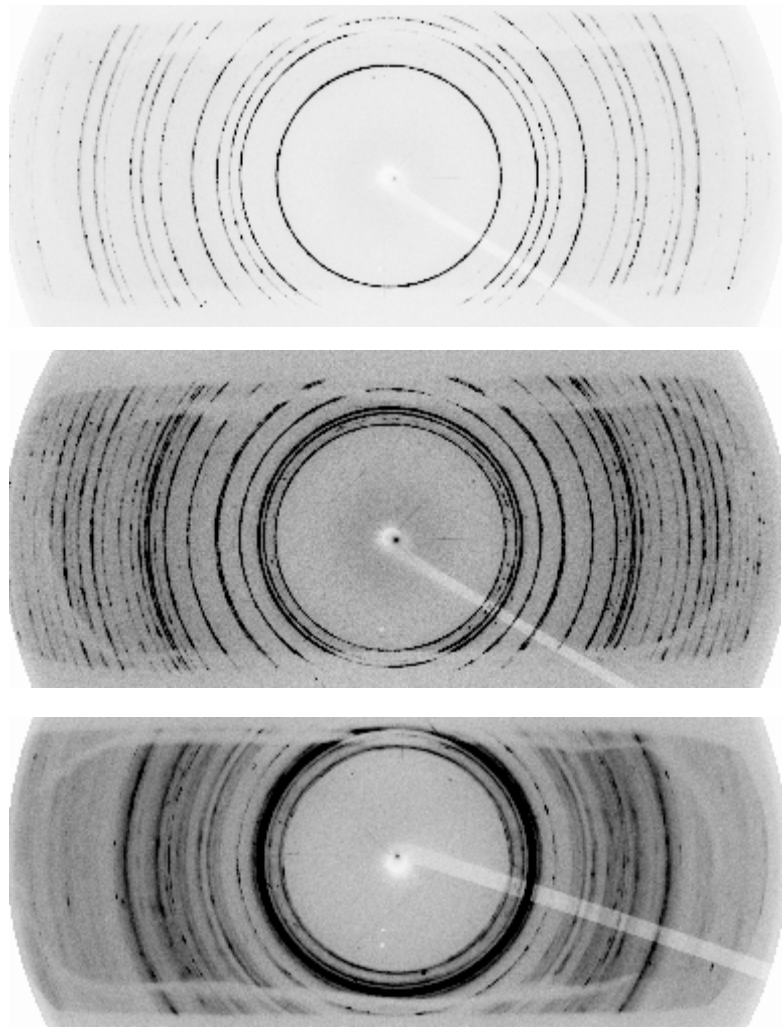

**Fig. 1.** X-ray diffraction images of the Eu sample and Pt pressure marker at 4 GPa (top, bcc phase), 14 GPa (center, hcp phase) with λ = 0.41493 Å beam and 2 s exposure time, and at 92 GPa (bottom, orthorhombic phase) with λ = 0.36229 Å beam and 15 s exposure time.

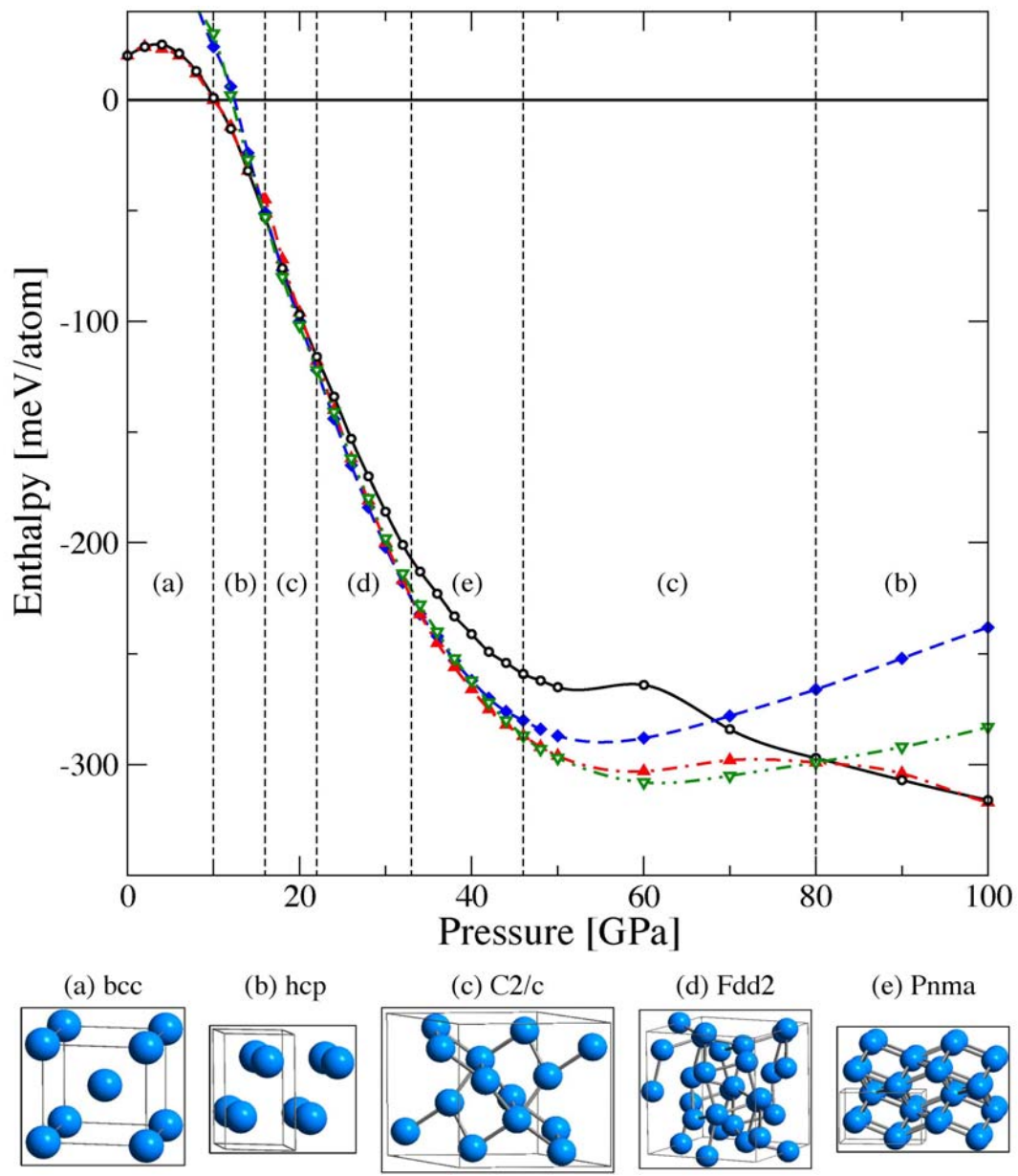

**Fig. 2.** Results of the random structure and genetic algorithm search at Cornell University showing the enthalpies of possible crystal structures of Eu relative to the bcc phase as a function of pressure up to 100 GPa. The DFT calculation predicts a structure sequence from bcc→hcp→*C*2/*c*→*Fdd*2→*Pnma*→*C*2/*c*→hcp. Figure legend: (a) bcc (horizontal line), (b) hcp (open circle), (c) *C*2/*c* (open triangle), (d) *Fdd*2 (diamond), (e) *Pnma* (solid triangle).

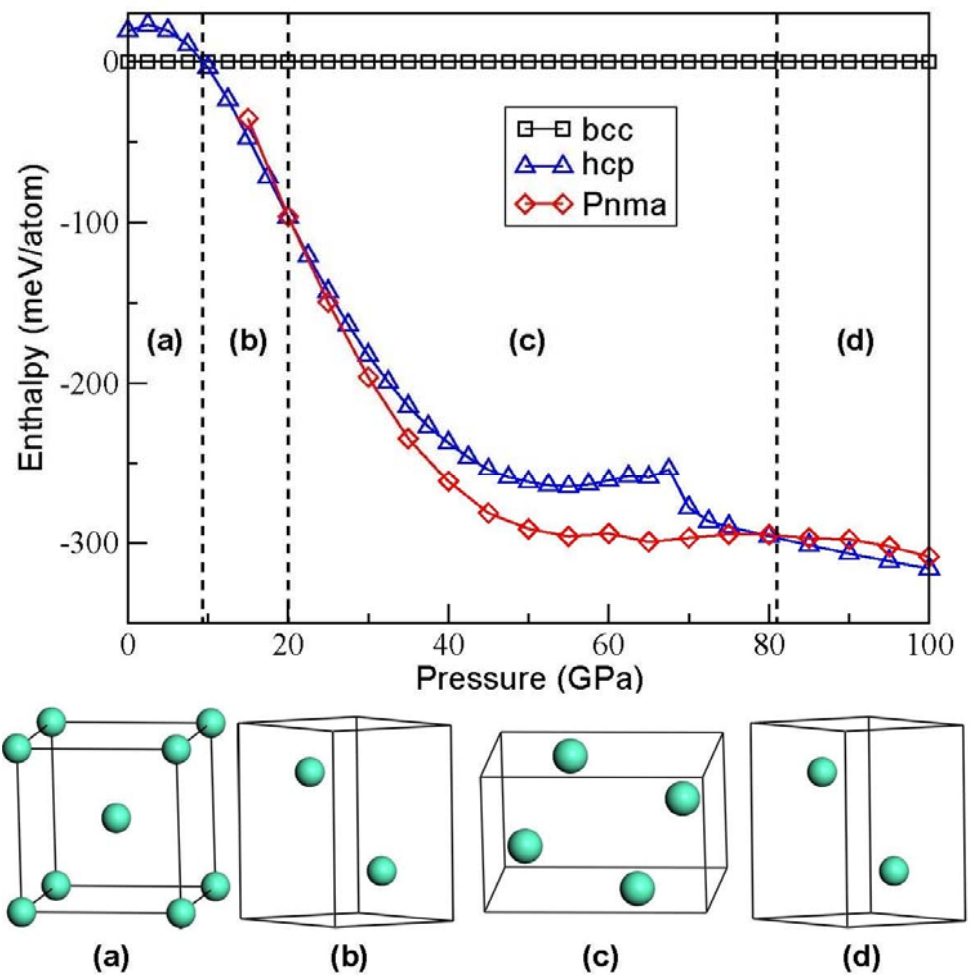

**Fig. 3.** Results of density function theory calculations at University of Nevada showing the enthalpies of possible crystal structures of Eu metal relative to that for the bcc phase as a function of pressure to 100 GPa: (a) bcc (square), (b) hcp *P6₃/mmc* (triangle), (c) orthorhombic *Pnma* (diamond), (d) hcp *P6₃/mmc* (triangle).

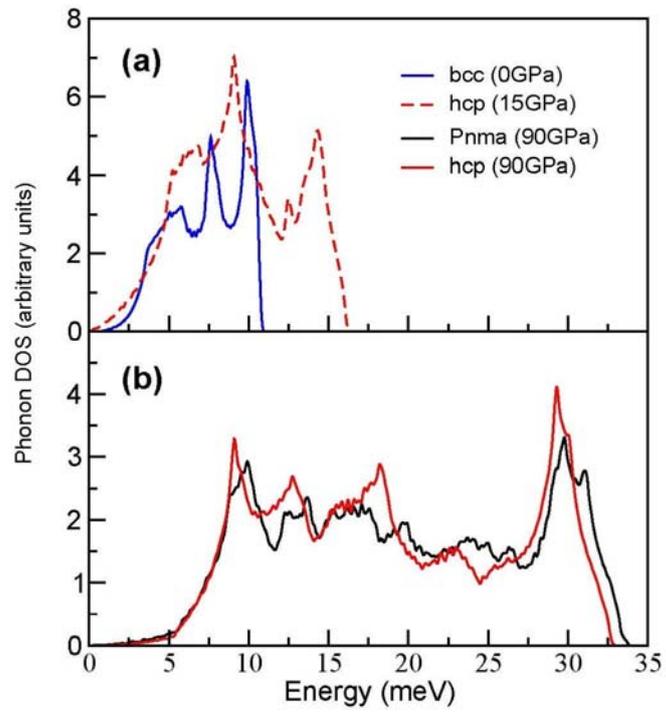

**Fig. 4.** Density of phonon states of Eu versus energy for structures in Fig. 3 at different pressures.

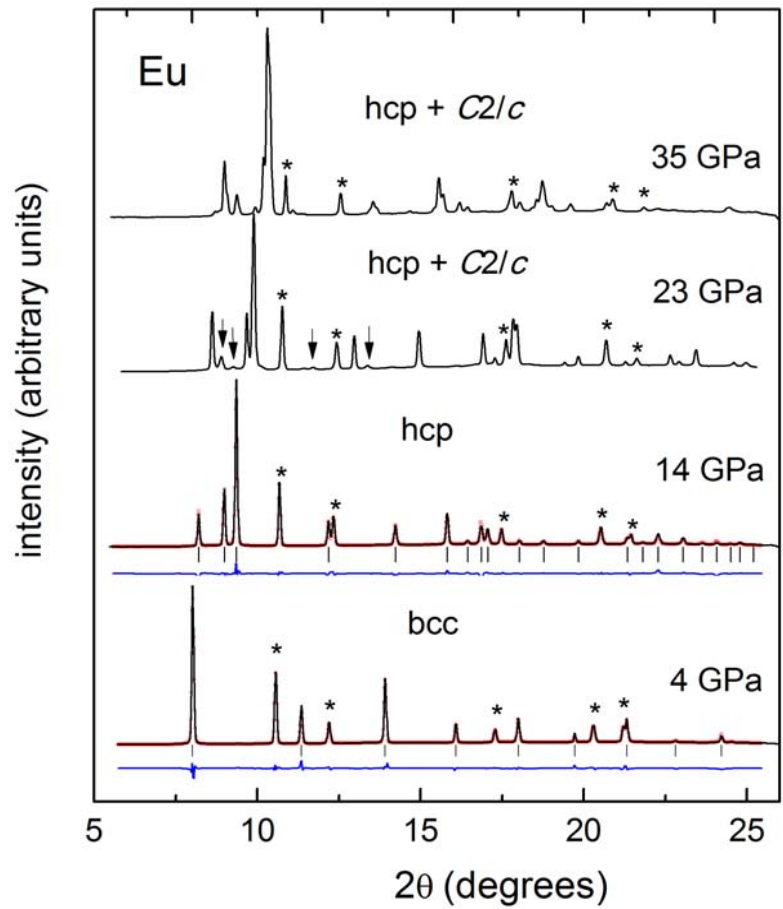

**Fig. 5.** Representative high pressure x-ray diffraction spectra of Eu (black lines, wavelength λ = 0.41493 Å) from 4 to 35 GPa with Rietveld full-profile refinements (red lines) for bcc and hcp phases. The tickmarks in the 4 GPa and 14 GPa plots correspond to positions of diffraction peaks of Eu. Below the tickmarks are the difference plots between calculated and observed spectra. Pt peaks are identified by asterisks in all spectra.

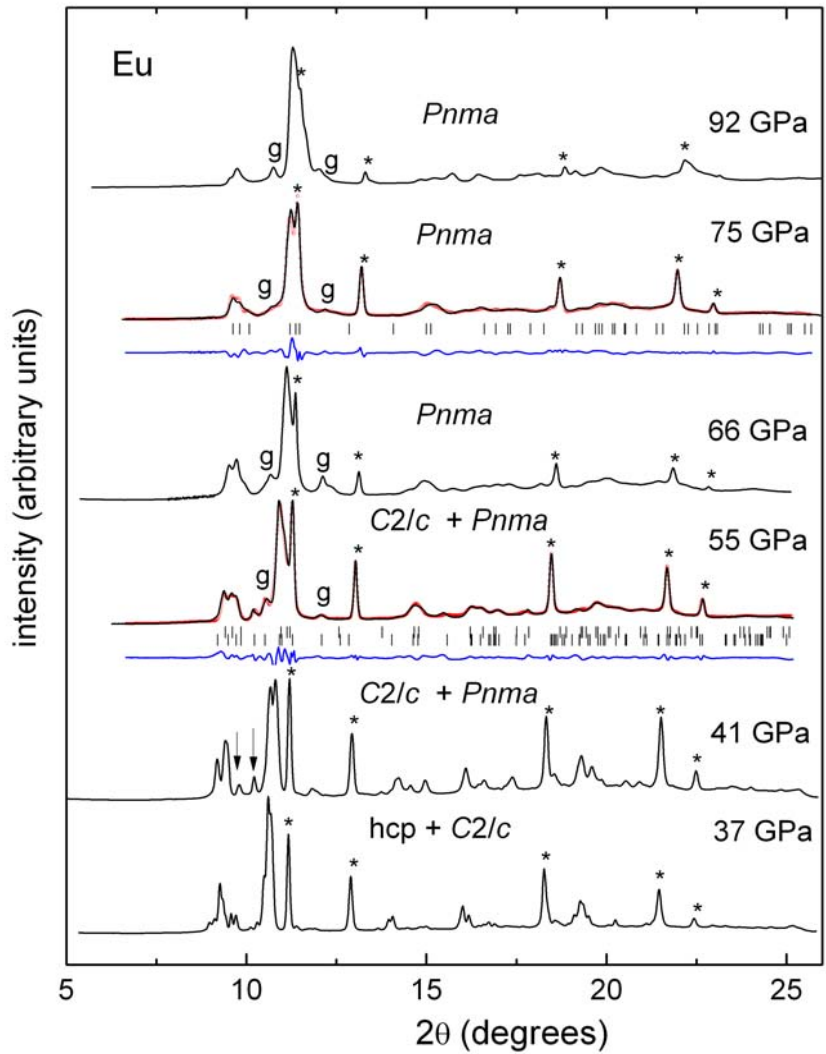

**Fig. 6.** Selected x-ray diffraction spectra of Eu (black lines, wavelength λ = 0.4256 Å) including the refinements (red lines) at 55 and 75 GPa showing the sluggish transition from *C*2/*c* to *Pnma*. In the plot for 55 GPa, the tickmarks correspond to the positions of diffraction peaks from Eu's *Pnma* phase (upper) and *C*2/*c* phase (lower). In the plot for 75 GPa, tickmarks show the peak positions from Eu's *Pnma* phase. The blue lines below the tickmarks show the difference plots between fits and data. Asterisks indicate peak positions from Pt. The letter "g" marks peaks from Re gasket.

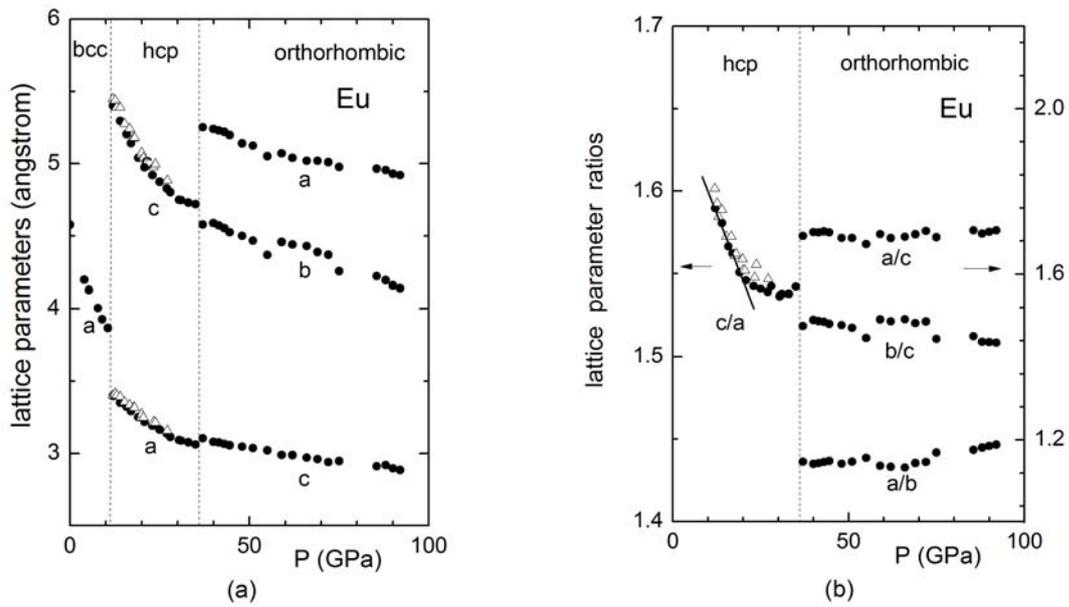

**Fig. 7.** For Eu pressure dependence of (a) lattice parameters and (b) ratio of lattice parameters above 12 GPa. In the pressure range 12 - 35 GPa, the lattice parameters are obtained from the hcp phase, while 35 - 92 GPa from the orthorhombic phase. The agreement of the *c/a* values from this study (solid circles) with those from Ref [5] (open triangles) is reasonable.

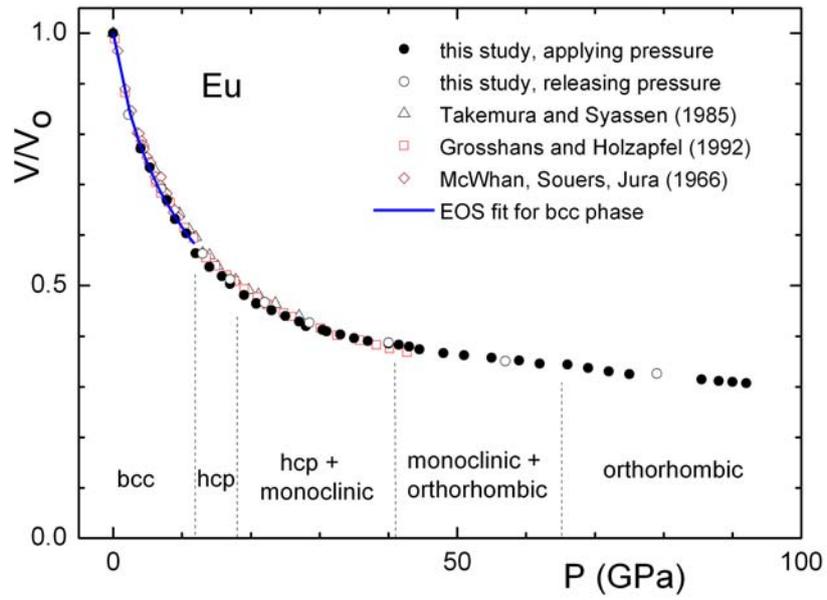

**Fig. 8.** Equation of state at ambient temperature for Eu to 92 GPa pressure from present studies compared to earlier work by Takemura and Syassen [5], Grosshans and Holzapfel [6], and McWhan, Souers, and Jura [29]. The *V(P)* fit in the bcc phase is obtained using the third order Birch-Murnaghan equation [30] (see text).

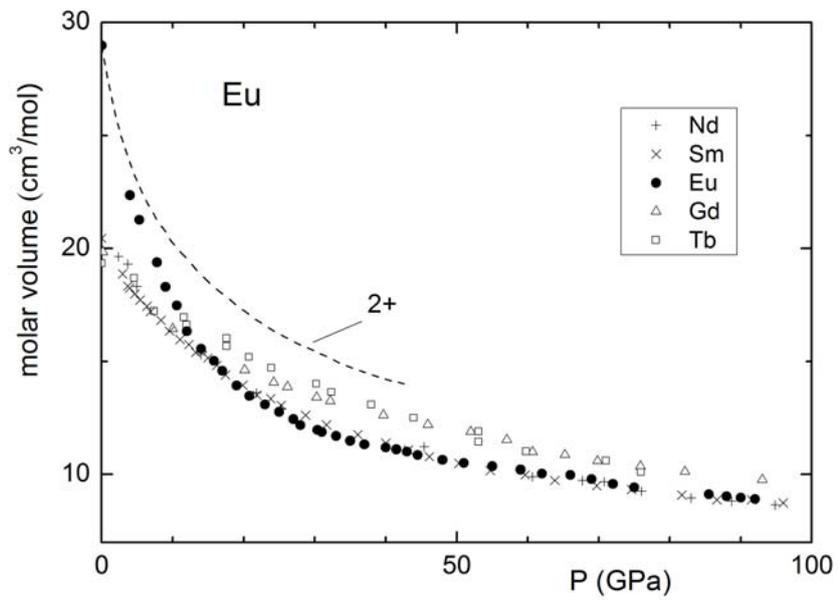

**Fig. 9.** Comparison of pressure-dependent molar volume of trivalent Nd [31], Sm [32], Gd [33], and Tb [34] to present results for Eu. Dashed line is calculation for divalent Eu by Johansson and Rosengren [2].